\documentclass[letterpaper, 10 pt, conference]{ieeeconf}
\IEEEoverridecommandlockouts
\usepackage{cite}
\usepackage{amsmath,amsfonts}
\usepackage{algorithmic}
\usepackage{graphicx}
\usepackage{textcomp}
\usepackage{xcolor}
\def\BibTeX{{\rm B\kern-.05em{\sc i\kern-.025em b}\kern-.08em
    T\kern-.1667em\lower.7ex\hbox{E}\kern-.125emX}}

\usepackage{bbm}
\usepackage{amsmath}

\usepackage{ntheorem}

\theorembodyfont{\rmfamily}

\newtheorem{Definition}{Definition} 

\newtheorem{Proposition}{Proposition} 

\newtheorem*{Proof}{Proof} 

\newtheorem{Assumption}{Assumption} 

\newtheorem{Remark}{Remark}

\newtheorem{Theorem}{Theorem}

\DeclareMathOperator{\col}{col}

\DeclareMathOperator{\diag}{diag}

\usepackage{amsmath}
\usepackage{amsfonts}
\usepackage{amssymb}
\usepackage{graphicx}
\usepackage{color}
\usepackage{float}
\usepackage{mathtools}

\usepackage{framed}
\usepackage{enumerate}

\DeclareMathAlphabet{\mymathbb}{U}{BOONDOX-ds}{m}{n}

\usepackage{array}

\usepackage{booktabs}

\usepackage{array}

\usepackage{balance}

\makeatletter
\renewcommand*\env@matrix[1][\arraystretch]{%
  \edef\arraystretch{#1}%
  \hskip -\arraycolsep
  \let\@ifnextchar\new@ifnextchar
  \array{*\c@MaxMatrixCols c}}
\makeatother

\usepackage{glossaries}
   
\newacronym{distributed generation unit}
{DGU}{distributed generation unit}

\newacronym{direct current}
{DC}{direct current}

\newacronym{Nash equilibrium problem}
{NEP}{Nash equilibrium problem}

\newacronym{point of common coupling}
{PCC}{point of common coupling}

\newacronym{generalized Nash equilibrium problem}
{DGU}{generalized Nash equilibrium problem}

\newacronym{generalized Nash equilibrium}
{GNE}{generalized Nash equilibrium}

\newacronym{normalized Nash equilibrium}
{NNE}{normalized Nash equilibrium}

\newacronym{distributed control and optimization framework}
{DCOF}{distributed control and optimization framework}

  \makeatletter
  \newcommand{\normal}{\bBigg@{0.5}}
\newcommand{\biggg}{\bBigg@{3}}

\newcommand{\Biggg}{\bBigg@{3.5}}

\newcommand{\bigggg}{\bBigg@{4}}

\newcommand{\Bigggg}{\bBigg@{4.5}}

\makeatother

\usepackage{stfloats} 
 
\usepackage[american,cute inductors,smartlabels]{circuitikz}
\usepackage{tikz}
\usetikzlibrary{arrows,automata}
\usetikzlibrary{arrows,shapes,backgrounds,calc,positioning,patterns}
\usetikzlibrary{calc}
\ctikzset{bipoles/thickness=1} 
\ctikzset{bipoles/length=1cm}
\tikzstyle{everynode}=[font=\small] \tikzstyle{every path}=[line width=0.8pt,line cap=round,line join=round]

\title{\LARGE \bf{A Distributed control framework\\ for the optimal operation of DC microgrids}}

\author{Zao Fu, Michele Cucuzzella, Carlo Cenedese, Wenwu Yu and Jacquelien M. A. Scherpen
\thanks{Corresponding author Jacquelien M. A. Scherpen, email: j.m.a.scherpen @rug.nl}
\thanks{Zao Fu and Wenwu Yu are with School of Cyber Science and Engineering, Southeast University, Nanjing, 211189, China.}
\thanks{Zao Fu and Jacquelien M. A. Scherpen are with Faculty of Science and Engineering, University of Groningen, Groningen 9747 AG, the Netherlands.}
\thanks{Michele Cucuzzella is with Department of Electrical, Computer and Biomedical Engineering, University of Pavia, Pavia 27100, Italy.}
\thanks{Carlo Cenedese is  with Automatic Control Laboratory, Department of Electrical Engineering and Information Technology,
        ETH Z\"urich, Physikstrasse 3 8092 Z\"urich, Switzerland. This research is supported   by NCCR Automation, a National Centre of Competence in Research, funded by the Swiss National Science Foundation (grant number 180545).} 
}

\begin{document}
\maketitle
\thispagestyle{empty}
\pagestyle{empty}

\begin{abstract}
In this paper we propose an original distributed control framework for DC mcirogrids.
We first formulate the (optimal) control objectives as an aggregative game suitable for the energy trading market.
Then, based on the dual theory, we analyze the equivalent distributed optimal condition for the proposed aggregative game and design a distributed control scheme to solve it.
By interconnecting the DC mcirogrid and the designed distributed control system in a power preserving way, we steer the DC microgrid's state to the desired optimal equilibrium, satisfying a predefined set of local and coupling constraints. 
Finally, based on the singular perturbation system theory, we analyze the convergence of the closed-loop system.
The simulation results show excellent performance of the proposed control framework.
\end{abstract}


\section{Introduction}
As an important part of the (actual and future) energy system, the \gls{direct current} microgrids are widely deployed in several applications, such as renewable energy sources, trains, aircraft, ships and charging stations for the more and more popular electric vehicles~\cite{cenedese:2019:PEV_charging}.
To improve the energy dispatch efficiency and trading fairness for the \gls{direct current} microgrids, one of the most effective options is the adoption of a \gls{distributed control and optimization framework}~\cite{YuW2017}.
Within such a framework, energy trading, control, and optimization processes will operate in a fully distributed way offering power stability, information privacy, plug-and-play capabilities, and market adaptability for large-scale power networks.
However, compared with the opponent centralized framework, the \gls{distributed control and optimization framework} requires to pay more attention to the design of the control objectives and deal with the constraints (especially with the coupling constraints).
In general, the control objectives for the DC microgrids focus on system-level requirements, e.g., system stability~\cite{CucuzzellaM2021} and convergence rate.
However, the optimization objectives might focus also on economic aspects, such as maximizing the profit from selling power, minimizing power costs, and reducing power losses.
For the control objectives, there are several results (see for instance \cite{CucuzzellaM2018,KosarajuKC2021} and the references therein).
We can divide the research on the optimization into two parts: modeling and algorithm design.
In the modeling part, convex optimization (e.g. quadratic programming) and non-cooperative games (e.g. aggregative games\cite{cenedese:2021:ad_geno},\cite{cenedese:2021:tv_prox_dyn}) are the most commonly used (see e.g.  \cite{Dorfler2017} and the references therein for further details).
In the algorithm design process, the challenges come from dealing with the following three aspects: local constraints, coupling constraints, and global information \cite{DePersisC2019}.
We mainly have three different methods for dealing with the local constraints.
The first one is called the \textit{penalty method}, and it uses a penalty function to embed the local constraints into the objective function \cite{FacchineiF2014}.
The second method is called the \textit{projection method}, and it restricts the descent direction within the feasible direction \cite{BelgioiosoG2018}.
The last method is the \textit{Lagrange multiplier method}, which dualizes the local constraints such that the corresponding dual problem does not have local constraints~\cite{boyd2004}.
On the other hand, one of the most effective methods for handling the coupling constraints is 
the \textit{multiplier consensus method}~\cite{YiP2019}.
Such a method employs the Lagrange dual method to first deal with the coupling constraints, and then converts the resulting dual problem into an optimization problem suitable for the design of distributed algorithms~\cite{YiP2019}.
For the aggregative information, one of the most common approaches is to design a (faster) estimation system (such as dynamical average consensus algorithms) to estimate the global information in a fully distributed fashion~\cite{LiangS2017}.

After modeling the control and optimization objectives and designing the control system, the next step is to connect the control system with the dynamics of the considered  \gls{direct current} microgrid.
Since the DC microgrid's dynamics can be shown to be passive, ensuring passivity of the controller as well, implies that, through  a suitable interconnection, the closed-loop system is still a passive system.
Inspired by such an idea, we design the control system based on the Lagrange dual theory, and prove that the closed-loop system converges to the desired (optimal) equilibrium, maximizing the profit while satisfying both the local and coupling constraints.

We organize the rest of the paper as follows.
Section \ref{Model description} introduces the dynamics of the DC microgrid and formulates the control and optimization objectives.
In Section \ref{problem analysis}, we analyze the distributed optimal condition, and 
in Section \ref{Algorithm design and analysis}, we design the distributed control scheme, interconnect it with the DC microgrid and analyze the closed-loop stability.
Section \ref{simulcation} shows the simulation results, and Section \ref{conclusion} concludes the paper.

\textbf{Notations}: 
$\col \left\{ x, y,\cdots \right\} =[ x^\top, y^\top,\cdots ] ^\top$, where the notation $\col$ represents ``vector stack".
Without additional explanation, we use $x$ to denote the `vector stack" of $x_1,\dots,x_n$, that is $x=\col\left\{ x _i \right\}_{i\in \mathcal{N}} =\col\left\{ x _1,\cdots ,x _n \right\}$.
The notation $\diag\normal\{x_i  \normal\}_{i \in \mathcal{N}}$ represents the (block) diagonal matrix whose diagonal entries consist of $x_1,\dots,x_n$.
Let  $\normal[ x \normal]_+=  \max\normal\{0,x \normal\}$.
The notation $\|x\|_A$, where $A$ is a symmetric definite or semi-definite matrix, represents the norm of semi-norm, and $\|x\|_A= \sqrt{\normal<x,Ax\normal>} $.
The notation $\delta_{\min}\normal( A \normal)$ denotes the minimal singular value of the matrix $A$.
The notation $\left\{x\right\}_i$ represents the $i$-th entry of the vector $x$.
The notation $\nabla$ denotes the gradient, and the notation $\nabla_x$ represents the gradient with respect to $x$.
The notation $\normal<x,y\normal>$ denotes the inner product of the vectors $x$ and $y$. 
The notations $\boldsymbol{0}_{n}$ and $\boldsymbol{1}_{n}$ denote $n$-dimension vectors whose entries are $0$ and $1$. 
Also, we omit the dimension when it is clear.
The notation $\partial_{x}  $ denotes the sub-gradient with respect to $x$.
The notation $J_{F,x}\left( x \right)$ denotes the Jacobin matrix of the function $F\left( x \right)$ with respect to $x$.
The notations $\circ$ and $\otimes$ represent the Kronecker product and the Hadamard product, respectively.
The symbols with the superscripts ``$r$" and ``$*$" denote constant references and  equilibriums (or Nash equilibriums).
The ``$\,\,\hat{}\,\,$" denotes the control states corresponding to the microgrid's ones. 

\section{Model description}
\label{Model description}
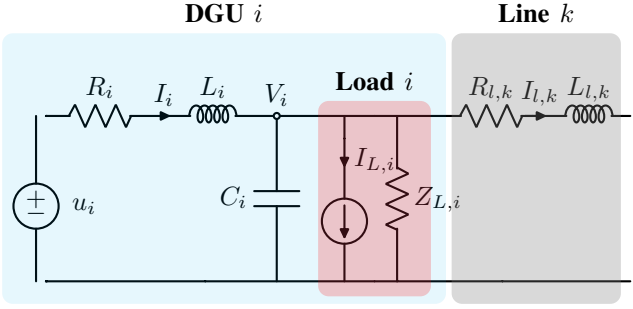
\begin{figure}
    	\begin{center}
    		\begin{circuitikz}
    			\ctikzset{current/distance=1}
    			\draw
    			node[] (Ti) at (0,0) {}
    			node[] (Tj) at ($(5.4,0)$) {}
    			node[] (Aibattery) at ([xshift=-4.5cm,yshift=0.9cm]Ti) {}
    			node[] (Bibattery) at ([xshift=-4.5cm,yshift=-0.9cm]Ti) {}
    			node[] (Ai) at ($(Aibattery)+(0,0.2)$) {}
    			node[] (Bi) at ($(Bibattery)+(0,-0.2)$) {}
    			(Ai) to [R, l={$R_{i}$}] ($(Ai)+(1.7,0)$) {}
    			($(Ai)+(1.7,0)$) to [short,i={$I_{i}$}]($(Ai)+(1.701,0)$){}
    			($(Ai)+(1.701,0)$) to [L, l={$L_{i}$}] ($(Ai)+(3,0)$){}
    			to [short, l={}]($(Ti)+(0,1.1)$){}
    			(Bi) to [short] ($(Ti)+(0,-1.1)$);
    			\draw
    			($(Ai)$) to []($(Aibattery)+(0,0)$)to [V=$u_{i}$]($(Bi)$)
    			($(Ti)+(-1.3,1.1)$) node[anchor=south]{{$V_{i}$}}
    			($(Ti)+(-1.3,1.1)$) node[ocirc](PCCi){}
    			($(Ti)+(-.4,1.1)$) to [short,i>={$I_{L,i}$}]($(Ti)+(-.4,0.5)$)to [I]($(Ti)+(-.4,-1.1)$)
    			
    			($(Ti)+(.3,1.1)$) to [R, l={$Z_{L,i}$}]($(Ti)+(.3,-1.1)$)to ($(Ti)+(.3,-1.1)$)
    			
    			($(Ti)+(-1.3,1.1)$) to [C, l_={$C_{i}$}] ($(Ti)+(-1.3,-1.1)$)
    			($(Ti)+(2.,1.1)$) to [short,i={$I_{l,k}$}] ($(Ti)+(2.2,1.1)$)
    			($(Ti)+(0,1.1)$)--($(Ti)+(.6,1.1)$) to [R, l={$R_{l,k}$}] 
    			($(Ti)+(2.5,1.1)$) {} to [L, l={{$L_{l,k}$}}, color=black]($(Tj)+(-2.2,1.1)$){}
    			($(Tj)+(-2.2,1.1)$) to [short]  ($(Ti)+(3.4,1.1)$)
    			($(Ti)+(0,-1.1)$) to [short] ($(Ti)+(3.4,-1.1)$);
    			\draw
    			node [rectangle,draw=none,minimum width=5.9cm,minimum height=3.6cm,dashed,fill=cyan,opacity=0.1,label=\textbf{DGU $i$},densely dashed, rounded corners] (DGUi) at ($0.5*(Aibattery)+0.5*(Bibattery)+(2.5,0.4)$) {}
			node [rectangle,draw=none,minimum width=1.5cm,minimum height=2.6cm,fill=red,opacity=0.2,dashed,label=\textbf{Load $i$},densely dashed, rounded corners] (DGUi) at ($0.5*(Aibattery)+0.5*(Bibattery)+(4.5,0)$) {}
			node [rectangle,draw=none,minimum width=2.25cm,minimum height=3.6cm,fill=gray,opacity=0.3,label=\textbf{Line $k$}, rounded corners] (DGUi) at ($0.5*(Aibattery)+0.5*(Bibattery)+(6.65,0.4)$) {};
    		\end{circuitikz}
    		\caption{Electrical scheme of DGU $i \in \mathcal{N}$ and transmission line $k \in \mathcal{E}$.}
    		\label{Equivalent circuit system for each DGU}
    	\end{center}
    \end{figure}

Following \cite{CucuzzellaM2021} and the references therein, we consider a microgrid consisting of a certain number of \glspl{distributed generation unit}, equipped with distributed controllers and decision systems.
Moreover, we consider that each \gls{distributed generation unit} includes 
constant impedance and constant current loads, and \glspl{distributed generation unit} are interconnected with each other via distribution power lines.
Let the sets $\mathcal{N}\triangleq \{1,\dots,n\}$ and $\mathcal{E} \triangleq \left\{ 1,\dots ,m \right\}$ denote the \gls{distributed generation unit} and the transmission line index sets, respectively.
For the readers’ convenience, Figure \ref{Equivalent circuit system for each DGU} shows the electric scheme of the \gls{distributed generation unit} $i$ and line $k$ (see also Table \ref{Physical description for the notations in Fig 1} for the description of the used symbols).


\begin{table}
\footnotesize
\centering
\caption{Physical descriptions for the notations.}
\begin{tabular}{lclc}
\toprule
\toprule
  Description & Symbol & Description & Symbol \\
\midrule
\midrule
Control input	&  $u_i$ & Filter parameters & $R_i, L_i$\\
Generated current	&  $I_i$ & Line parameters & $R_{l,k}, L_{l,k}$\\
Line current	&  $I_{l,i}$	 & Shunt capacitor & $C_i$\\
Load voltage	&  $V_i$	&Load parameters & $I_{L,i}, Z_{L,i}$\\
\bottomrule
\bottomrule
\end{tabular}
\label{Physical description for the notations in Fig 1}
\end{table}
According to Kirchhoff's law, we can write the dynamics of the DC microgrid as follows (see e.g.  \cite{CucuzzellaM2021}):
\begin{align}
\begin{split}
L\dot{I}=&-V-RI+u, \\
C\dot{V}=&~I+BI_l-Z_L^{-1}V-I_L\\
L_l\dot{I}_l=&-R_lI_l-B^\top V.
\end{split}
\label{compact dynamical system}
\end{align}
where $B\in \mathbb{R}^{m\times n}$ is the adjacency matrix associated with the arbitrary oriented graph of the connected and undirected graph $\mathcal{G}\left(\mathcal{N},\mathcal{E}\right)$.  
We assume that each transmission line is under the control of a unique \gls{distributed generation unit} connected to it, and we let $\mathcal{E}_i$ represent the index set of the  transmission lines being under the control of the \gls{distributed generation unit} $i\in \mathcal{N}$.
Hence, we have
\begin{align}
		 \bigcap_{i\in \mathcal{N}}{\mathcal{E}_i=\emptyset}, \,\,\,\,\,		 \bigcup_{i\in \mathcal{N}}{\mathcal{E}_i}=\mathcal{E}.
\label{definition of C}
\end{align}
Let now define for convenience $x_i \triangleq \col \{ I_i, V_i, I_{c,i}\} \in \mathbb{R}^{2+\left| \mathcal{E}_i \right|}$, with $I_{c,i}\triangleq \col\{ I_{l,k} \}_{k\in\mathcal{E}_i} \in \mathbb{R}^{\left| \mathcal{E}_i \right|}$,
to denote the state vector of the DGU $i\in \mathcal{N}$.
The objective of this paper is to design a distributed control scheme that stabilizes the considered DC microgrid at the desired equilibrium solving a pre-designed game problem.
To achieve this goal, we use passivity theory \cite{Van2000} to interconnect the considered microgrid's dynamics \eqref{compact dynamical system} with the distributed control system we design in the following sections.

Before formulating the game problem, we describe the feasible region of operation of the considered microgrid by  introducing the following set of coupling constraints:
\begin{align}
K \triangleq&~\Bigg\{ \normal(u, x\normal) \in \mathbb{R}^{m+3n} \,             \Bigg| \, \underset{Ax-s_A~=~\boldsymbol{0}_{m+n}}{\underbrace{\begin{matrix}[1.5]
	I+B I_l-Z_{L}^{-1}V-I_L=\boldsymbol{0}_n\\
	R_{l} I_l+B^{\top}V=\boldsymbol{0}_m\\
\end{matrix}}}  \Bigg\}, 
\label{global constraint}
\end{align}
where we omit the definitions of the matrix $A \in \mathbb{R}^{\left(m+n\right)\times n}$ and the vector $b\in \mathbb{R}^{m+n}$.
Moreover, for every DGU $i\in \mathcal{N}$, we introduce the following set of local constraints: 
\begin{align}
\Omega_i \triangleq \left\{ \left( u_i, x_i \right) \in \mathbb{R}^{3+\left| \mathcal{E} _i \right|} \left|\,\, \begin{matrix}[1.5]
	V_i+R_iI_i-u_i=0\\
	V_{i}^{\min}\le V_i\le V_i^{\max}\\
	 I_{c,i}^{\min}\le I_{c,i} \le I_{c,i}^{\max}
\end{matrix} \right. \right\},
\label{local constraint}
\end{align}
where the superscripts `$\min$' and `$\max$' represent the minimum and maximum values of the the corresponding state, respectively.
Based on  \eqref{global constraint} and \eqref{local constraint}, we can then define the following feasible set:
\begin{align}
&\pi_i \left( x_{-i} \right) =\nonumber\\
&~\Bigg\{ \normal( u_i, x_i \normal) \in \Omega _i\mid\underset{\phi_i\left(x_i\right)}{\underbrace{ A_ix_i+s_{A_i}}}+
\underset{\phi_{-i}\left(x_{-i}\right)}{\underbrace{\sum_{j=1,j\ne i}^n{A_jx_j+s_{A_j}}}}= \boldsymbol{0}\Bigg\}, 
\label{coupling constraint + local constraint}
\end{align}
where for all $i \in \mathcal{N}$,  $A_i \in \mathbb{R}^{\left( m+n \right) \times \left( 3+\left| \mathcal{E}_i \right|\right)}$ and $s_{A,i}$ are  constant  and  satisfy $A=\left[ A_1,\dots ,A_n \right], \sum_{i=1}^{n}{s_{A_i}}=s_A$.
Moreover, $x_{-i}$ denotes the stack of all the rivals' decisions, i.e., $x_{-i}=\col \left\{x_1,\dots,x_{i-1},x_{i+1},\dots,x_n \right\}$. 

We can now introduce the main goal of the paper, which can be formulated as an aggregative game probelem, i.e.,  for all $i \in \mathcal{N}$
\begin{align}
\begin{split}
\begin{matrix}[2]
 \min\limits_{u_i, x_i} & f_i\left( u_i, x_i, x_{-i} \right)\\ 
\mathrm{s.t.} & \normal( u_i, x_i \normal
)\in \pi_i \left( x_{-i} \right),
\end{matrix}
\end{split}
\label{primal game problem}
\end{align}
with
\begin{align}
   \begin{split}
   f_i\left( u_i, x_i, x_{-i} \right) &= f_{1,i}\left( u_i, x_i \right) + f_{2,i}\left( x_i,x_{-i} \right)\\
  f_{1,i}\left( u_i, x_i \right) & \triangleq {{\frac{\alpha _{u_i}}{2} \left( u_i-u_{i}^{r}\right)^{2}+\frac{1}{2}\|x_i-x_{i}^{r}\|_{A_{x_i}}^{2}}}\\
f_{2,i}\left( x_i,x_{-i} \right) & \triangleq - {{\Big(l-p_r\underset{s_{p_r}}{\underbrace{\sum\nolimits_{i=1}^n{I_i}}}\Big)V_i^{r} I_i}},\\
   \end{split}
\end{align}
where $f_{1,i}( u_i, x_i)$ represents the cost associated with the deviation of the $i$-th DGU's state and input with respect to the corresponding references, while $f_{2,i}( x_i,x_{-i})$ represents the profit of DGU $i$, where $(l - p_r  s_{{p_r}})>0$ is the selling price of the generated power $V_i^r I_i$. Moreover, $\alpha_{u_i}$ is a positive constant and the matrix ${A_{x_i}} \triangleq \diag\left\{ \alpha_{I_i}, \alpha_{V_i},\diag\big\{\alpha _{I_{l,k}}\normal\}_{ k\in \mathcal{E}_i}  \right\}$ 
is positive definite;
the parameters $p_r$ and $l$ are positive constants {{ensuring that the price of power is always positive, i.e., $(l - p_r  s_{{p_r}})>0$}} for all the feasible currents $I_1,\dots, I_n $.
According to the constraint~\eqref{global constraint},
we can guarantee such a condition by introducing the following assumption:
\begin{Assumption} (\textbf{Parameter setting)}
Let the following condition
\begin{align}
0<l-p_r \sum_{i=1}^n{\left( \frac{V_i^{\max}}{  Z_{L,i}}+I_{L,i} \right)}
\label{How to set pr and l}
\end{align}
hold for all $i\in \mathcal{N}$.
\end{Assumption}
Note that the increase of the generated currents' sum $s_{p_r}$ implies a reduction of the power price $l-p_r s_{p_r}$ (and vice versa), as usual in the energy trading market.

\section{Problem analysis}
\label{problem analysis}
In this section, we analyze the optimality conditions associated with the game problem \eqref{primal game problem}.
First, we introduce the definition of the \gls{generalized Nash equilibrium} \cite[Definition 1]{DePersisC2019}.
\begin{Definition} (\textbf{Generalized Nash equilibrium})
The point $\left( u^*, x^* \right)$ is a \gls{generalized Nash equilibrium} for the game \eqref{primal game problem} if and only if it solves the following problem: 
\begin{align}
\min_{u_i, x_i}  \,\,\,\,\,		f_i \normal( u_i, x_i, x_{-i}^{*} \normal) \,\,\,\,\,			\textrm{s.t.} \,\,\,\,\,			\normal( u_i, x_i \normal) \in \pi_i \normal( x_{-i}^{*} \normal).
\label{equilent problem}
\end{align}
 for all $i \in \mathcal{N}$.
\end{Definition}
Now, before formulating the dual problem of \eqref{equilent problem},  
for all $i\in \mathcal{N}$, we define the  following penalty function (distance function):
\begin{align*}
  g_{i}\normal( x_i  \normal)=&~\underset{=~g_{1,i}\normal( V_i  \normal)}{\underbrace{\rho_{V_i}  \big(\normal[ V_i^{\min}-V_i \normal]_+ + \normal[ V_i-V_i^{\max} \normal]_+ \big)}} 
   \\
  & +\sum\nolimits_{k\in \mathcal{E}_i }\underset{=~g_{2,k}\normal( I_{l,k}  \normal)}{\underbrace{  \rho_{I_{l,k}}  \big(\normal[ I_{l,k}^{\min}-I_{l,k} \normal]_+ + \normal[ I_{l,k}-I_{l,k}^{\max} \normal]_+ \big ) }},
\end{align*}
where the positive constants $\rho_{V_i}$ and $\rho_{I_{l,k}}$, $k\in \mathcal{E}_i$, represent the penalty parameters.
According to \cite[Lemma 4]{FacchineiF2014}, there exist positive penalty parameters for the sub-problems in \eqref{equilent problem} such that its solution and the solution to the corresponding penalized problem coincide.
For all $i\in \mathcal{N}$, we can then define the following Lagrange function
\begin{align}
\begin{split}
 	\mathcal{L}_i \normal( u_i, x_i, x_{-i}^{*},\gamma_i,\lambda _i \normal) =&~f_i \normal( u_i, x_i, x_{-i}^{*} \normal)\\
 	&+g_{i}\normal( x_i  \normal)+\gamma_i \normal(D_i^\top x_i-u_i\normal)\\
&+\normal< \lambda _{i},\phi_i \normal( x_i\normal)+\phi_{-i} \normal( x_{-i}^{*} \normal) \normal>,
\end{split}
\label{local Lagrange function}
\end{align}
where $D_i= \col\normal\{1,R_i, \boldsymbol{0} \normal\} \in \mathbb{R}^{2+\left| \mathcal{E}_i \right|}$ is a  constant vector
and $\gamma_i \in \mathbb{R}$ represents the  Lagrange multipliers associated with the constraints $D_i^\top x_i-u_i=0$ (see the first equality constraints in \eqref{local constraint}). Similarly, $\lambda_i \in \mathbb{R}^{m+n}$ represents the  Lagrange multipliers associated with the coupling constraints in \eqref{coupling constraint + local constraint}.
Note also that the penalty function $g_i \normal( x_i \normal)$ is not differentiable at some points.
Hence, we need to introduce the sub-gradient of $g_i \normal( x_i \normal)$ consisting of the sub-gradients of the penalty functions $g_{1,i} \normal( V_i  \normal)$ and $g_{2,k}\normal( I_{l,k}  \normal)$.
For all $i\in \mathcal{N}$, the sub-gradient of $g_{1,i} \normal( V_i  \normal)$ is as follows \cite{Cherukuri2015}
\begin{align}
    \partial g_{1,i} \normal( V_i  \normal) \in \left\{ 
    \begin{matrix}[1.3]
    -\rho_{V_i} & \mathrm{if} \,\,      V_i < V_i^{\min},  \,\,\,\,\,\,\,\,\,\,\,\,\,\,\,\,\,\,\,\,    \\
     [-\rho_{V_i}, 0 ] & \mathrm{if} \,\,V_i = V_i^{\min}, \,\,\,\,\,\,\,\,\,\,\,\,\,\,\,\,\,\,\,\,  \\
     0 & \mathrm{if} \,\,V_i^{\min}< V_i < V_i^{\max}, \\
     [0, \rho_{V_i} ] & \mathrm{if} \,\,V_i = V_i^{\max}, \,\,\,\,\,\,\,\,\,\,\,\,\,\,\,\,\,\,\,\,  \\
     \rho_{V_i} & \mathrm{if} \,\,      V_i > V_i^{\max}. \,\,\,\,\,\,\,\,\,\,\,\,\,\,\,\,\,\,\,\, \\
    \end{matrix}
    \right.
    \label{subgradient}
\end{align}
The sub-gradient of  $g_{2,i} \normal( I_{l,k}  \normal) $ can be obtained as in \eqref{subgradient}, thus we omit it for the sake of simplicity.
To guarantee that a GNE exists for the game problem \eqref{primal game problem}, we introduce the following assumption:
\begin{Assumption}(\textbf{Non-empty feasible set})
The feasible set given by the intersection of $K$ in \eqref{global constraint} and $\Omega_1,\dots, \Omega_n$ in \eqref{local constraint} 
 is non-empty.
\label{non-empty assumption}
\end{Assumption}
Since the constraints of each sub-problem in \eqref{primal game problem} are affine, Assumption \ref{non-empty assumption} guarantees that $\pi_i$ satisfies Slater's constraint qualification. Therefore, the following KKT conditions are a necessary and sufficient condition for the optimal condition of the problem \eqref{primal game problem} (refer to   \cite[Section 5.2.3]{boyd2004} for details):
\begin{align}
\forall\, i \in \mathcal{N},\,\, \left\{ \begin{matrix}[1.5]
	\nabla_{u_i} \mathcal{L}_i \normal( u_i^*, x_i^*, x_{-i}^*,\gamma_i^*,\lambda _i^* \normal) =\boldsymbol{0}_{n},\,\,\,\,\,\,\,\,\,\,\\
	\partial_{x_i} \mathcal{L}_i \normal( u_i^*, x_i^*, x_{-i}^*,\gamma_i^*,\lambda _i^* \normal) \ni \boldsymbol{0}_{m+2n},\\	
	\nabla_{\lambda_i} \mathcal{L}_i \normal( u_i^*, x_i^*, x_{-i}^*,\gamma_i^*,\lambda _i^* \normal)=\boldsymbol{0}_{m+n},\\
	\nabla_{\gamma_i} \mathcal{L}_i \normal( u_i^*, x_i^*, x_{-i}^*,\gamma_i^*,\lambda _i^* \normal)=0,\,\,\,\,\,\,\,\,\,\,
\end{matrix} \right.
\label{KKT condition for first dual}
\end{align}
where $\normal( u_i^*, x_i^*,\gamma_i^*,\lambda _i^* \normal)$ is the saddle-point of the Lagrange function \eqref{local Lagrange function}.
Since the multipliers $\lambda_1^*,\dots,\lambda_n^*$ can vary from each other, the solution of the KKT condition \eqref{KKT condition for first dual} may not be unique.
To shrink the solution set of the KKT condition \eqref{KKT condition for first dual} to a convex set (or a singleton), 
such that we can develop a fully distributed algorithm, we need to introduce the following definition \cite[Definition 3.2]{NabetaniK2021},\cite{cenedese:2020:time_var_constrained}.

\begin{Definition}(\textbf{Normalized Nash equilibrium})\label{Normalized Nash}
A GNE $\normal(u^*,$ $x^* \normal)$ is a \gls{normalized Nash equilibrium} associated with the given $r_1,\dots,r_n >0$, if there exist the Lagrange multipliers $\gamma^*$ and $\lambda^*$  such that $\normal(u^*, x^*, \gamma^*, \lambda^* \normal)$ solves the KKT condition \eqref{KKT condition for first dual} and satisfies the additional condition
\begin{align}
r_1\lambda _{1}^{*}=\cdots =r_n\lambda_n^{*}.
\label{multipliers consensus constraints}
\end{align}
\end{Definition}
\begin{Remark}
The values of the Lagrange multipliers $\lambda_1^*, \dots, $ $\lambda_n^* $ concerning the coupling constraints represent the shadow price of all the \glspl{distributed generation unit}. 
From a trading market point of view, the values of $r_1, \dots, r_n$ can be designed   by a higher level decision system (for example, the government) in order to model different market scenarios.
\end{Remark}

For the sake of analysis, let  $A_r\triangleq \diag\normal\{r_i  \normal\}_{i \in \mathcal{N}} \otimes \mathbb{I}_{m+n},  \lambda \triangleq \col\normal\{ \lambda_i \normal\}_{i \in \mathcal{N}}, \boldsymbol{L} \triangleq 
\mathbb{L} \otimes \mathbb{I}_{m+n}$, 
where $\mathbb{L}$ represents the Laplacian matrix associated with $\mathcal{G}$.
Since $\mathcal{G}$ is undirected and connected, then the condition \eqref{multipliers consensus constraints} is equivalent to 
$
\boldsymbol{L} A_r \lambda=\boldsymbol{0}_{n(m+n)}
$.
Then, we introduce the following proposition playing a crucial role in the later controller design, as in \cite{cenedese:2021:ad_geno}. 

\begin{Proposition}
\label{consensus constraint equivalence}
There exist $\upsilon^* \triangleq \col \normal \{ \upsilon_i^* \normal \}_{i\in \mathcal{N}}$, $\nu^* \triangleq \col\normal \{ \nu_i^* \normal \} _{i\in \mathcal{N}} \in \mathbb{R}^{n}$
 satisfying
\begin{align}
 \left\{ \begin{matrix}[1.2]
\displaystyle  -\normal( \mathbb{I}+\mathbb{L} \normal)  	\upsilon^* - \mathbb{L} \nu ^* +nI^*= \boldsymbol{0}_n, \\
	\mathbb{L}  \upsilon ^* =\boldsymbol{0}_n\\
	\end{matrix} \right.
	\label{KKT condition for fast system}
\end{align}
if and only if
\begin{align}
\upsilon_1^*=\cdots=\upsilon_n^*=\sum_{i=1}^{n}{I_i^{\,*}},
\label{optimal condition for fast system}
\end{align}
where $I_i^* \in \mathbb{R}$ for all $i \in \mathcal{N}$.
\end{Proposition}
\begin{Proof}
It holds that
\begin{align}
	\mathbb{L}  \upsilon ^* =\boldsymbol{0}_n\,\,\,\Leftrightarrow \,\,\, \upsilon_1^* =\cdots=\upsilon_n^*.
\label{equivalence of Laplacian matrix and consensus constraint}
\end{align}
By substituting the second equality of \eqref{KKT condition for fast system} in the first equality and multiplying both sides by $\boldsymbol{1}_n^\top $, we can obtain the condition \eqref{optimal condition for fast system}.
From \eqref{optimal condition for fast system} and $\mathrm{rank}(\mathbb{L}) = n-1$, we  deduce that there exists $\nu^*$ satisfying
\begin{align}
\upsilon^*- nI^*= \Big( \sum_{i=1}^{n}{I_i^*} \Big)\, \boldsymbol{1}_n - nI^*=\mathbb{L}  \nu ^*.
\label{G1_i}
\end{align}
By combining \eqref{G1_i} and \eqref{equivalence of Laplacian matrix and consensus constraint}, we obtain the condition \eqref{KKT condition for fast system}, which completes the proof. \hfill \QED
\end{Proof}
According to \cite[Proposition 3.2]{NabetaniK2021}, the \gls{normalized Nash equilibrium} associated with a given $r>0$ of the problem \eqref{primal game problem} corresponds to the solution of the following variation inequality:
\begin{align}
x^* \in K \cap \Omega,\,\, \,\,  \normal< F_r \normal( u
^*, x^* \normal
), x-x^* \normal> \ge 0, \,\,\,\,  \forall\, x \in K \cap \Omega,
\label{variational inequality}
\end{align}
where $ \Omega \triangleq \bigcap\nolimits_{i=1}^n {\Omega_i}$, and the vector function $F_r\normal( u, x \normal)$ is the pseudo-gradient (refer to \cite{NabetaniK2021}) defined as follows
\begin{align*}
F_r \normal( u, x \normal)= \col\normal\{ r_i \nabla_{(u_i, x_i)} f_i \normal( u_i, x_i, x_{-i} \normal) \normal\}_{i\in \mathcal{N}}.
\end{align*}
To ensure that the variational inequality \eqref{variational inequality} (as well as the problem \eqref{primal game problem}) has a unique \gls{normalized Nash equilibrium} $(u^\ast, x^\ast)$ for a fixed $r>0$, we need to introduce the following assumption:
\begin{Assumption}\label{Monotonicity Assumption}(\textbf{Bound for parameters})
For all $i\in\mathcal{N}$, the parameter $r_i$ satisfies  the following condition:
\begin{align}\label{eq:mon_cond_eq}
2r_i\alpha _{I_i}+\left( 6-n \right) r_ip_rV_{i}^{r}-\sum_{i=1}^n{r_ip_rV_{i}^{r}}>0.
\end{align}
\end{Assumption}
Under Assumption \ref{Monotonicity Assumption}, one can verify that the Jacobian matrix $ J_{F_r} \normal( u, x \normal) \succ 0$ and thus $F_r(u,x)$ is strict monotone for all $(u,x) \in K \cap \Omega$ (refer to  \cite[Theorem 2.3.3]{FacchineiF2003}).
Therefore, the variational inequality \eqref{variational inequality} has a unique solution under Assumptions   \ref{non-empty assumption} and \ref{Monotonicity Assumption}.
All the parameters in \eqref{eq:mon_cond_eq} have to be designed, and thus Assumption \ref{Monotonicity Assumption} is not a strict condition.
Following from the analysis in Proposition \ref{consensus constraint equivalence}, we can deduce that the following constraint 
\begin{align*}
\left\{
\begin{matrix}[1.3]
\displaystyle   A_r \normal( \col \normal\{ A_i x_i^* -s_{A_i} \normal\}_{i\in \mathcal{N}} - \boldsymbol{L} A_r \lambda ^* 
- \boldsymbol{L} \theta ^* \normal) =\boldsymbol {0}_{n\left(m+n\right)},
\\
	\displaystyle \boldsymbol{L} A_r\lambda ^* 
=\boldsymbol {0}_{n\left(m+n\right)}\\
\end{matrix}\right.
\end{align*}
is equivalent to the constraint $Ax^*-s_A=\boldsymbol{0}_{m+n}$.
Therefore, by involving the constraint \eqref{multipliers consensus constraints} and Proposition \ref{consensus constraint equivalence}, we can rewrite the condition \eqref{KKT condition for first dual} is a distributed form as 
\begin{align}
\forall\, i\in \mathcal{N},\,\left\{ \, \begin{matrix}[1.3]
\displaystyle 	-\upsilon _i^*  -\mathbb{L}_i \upsilon ^* 
-\mathbb{L}_i \nu ^* 
+nI_i^*=0,\\
	\displaystyle \mathbb{L}_i \upsilon ^*  =0,\\
	r_i\alpha _{u_i} \normal( u_i^*-u_i^r \normal)+\gamma_i^*=0,
	\\
	r_i \bar{F}_i \normal (\, x_{i}^{*},\upsilon_i^* \,\normal) +r_iA_{i}^\top \lambda _{i}^{*}+\gamma _{i}^{*} D_i \ni \boldsymbol{0}_{m+3n},\\
		D_i^\top x_i^*-u_i^*  =0,\\
	\displaystyle   r_i \normal( Ax_{i}^{*}-s_{A_i} \normal) -r_i \boldsymbol{L}_i A_r \lambda ^* 
-r_i \boldsymbol{L}_i \theta ^*  
= \boldsymbol{0}_{m+n},\\
	\boldsymbol{L}_i  A_r \lambda ^*
=\boldsymbol{0}_{m+n},\\
\end{matrix} \right.
\label{Final KKT condition} 
\end{align}
where for all $i\in \mathcal{N}$, the vector $\theta_i^* \in \mathbb{R}^{m+n}$ denotes the dual variables  associated with the consensus constraint $\boldsymbol{L}_i A_r  \lambda^* =\boldsymbol{0}_{m+n}$. We use the vector $\mathbb{L}^\top_i \in \mathbb{R}^n$ and matrix $\boldsymbol{L}_i \in \mathbb{R}^{\left( m+n \right
)\times n\left( m+n \right
)}$ to denote the rows of $\mathbb{L}$ and $\boldsymbol{L}$ associated with the DGU $i\in \mathcal{N}$, respectively.
Moreover, $\bar{F}_i \left (\hat{x}_{i}^*,\upsilon_i^* \right)$ is defined as
\begin{align*}
\bar{F}_i \left (\hat{x}_{i}^*,\upsilon_i^* \right) 
=\left[\, \begin{matrix}[1.5]
	\alpha _{I_i} \normal(  I_i^*-I_i^r \normal)-\left( l-p_rV_i^{r}\upsilon _i^* \right) +p_rV_i^{r}I_i^*\\
	\alpha _{V_i}\normal( V_i^*-V_i^r \normal) +\partial  g_{1,i}\normal( V_i^*  \normal) \\
	\col \normal\{ \normal( I_{l,k}^*-I_{l,k}^r \normal)+\partial  g_{2,k}\normal( I_{l,k}^*  \normal) \normal\}_{ k\in \mathcal{E}_i}\\
\end{matrix}\, \right].
\end{align*}

\begin{Remark}
Based on  \cite[Theorem 3.4]{HanSP1979}, the penalty parameters $\rho_{V_i}$ and $\rho_{I_{l,k}}$ satisfy the following condition
\begin{align}
 \rho_{V_{i}}\ge&~\alpha_{V_i} \normal( V_i^{\max}- V_i^r \normal)+\nabla_{V_i}\normal<\lambda_i^*, \normal( Ax-s_A \normal) \normal>+\gamma_i^*,
 \notag \\
  \rho_{I_{l,k}}\ge &~\alpha_{I_{l,k}} \normal(I_{l,k}^{\max}-  I_{l,k}^r \normal)+\nabla_{I_{l,k}}\normal<\lambda_i^*, \normal( Ax-s_A \normal) \normal>.
  \label{penalty parameter}
\end{align}
for all $i\in\mathcal{N}$ and $k\in \mathcal{E}_i$.
Hence, the penalty parameters should be large enough such that they satisfy  \eqref{penalty parameter}.
\end{Remark}

\section{Algorithm design and analysis}
\label{Algorithm design and analysis}
Based on  \eqref{Final KKT condition}, we can now design the distributed controller for each DGU $i\in \mathcal{N}$.
By connecting the designed controller to the DGU $i\in \mathcal{N}$ in a passive way (see e.g. \cite{CucuzzellaM2021}), we obtain the following closed-loop system:
\begin{subequations}
\label{whole sysem}
\begin{align}
\dot{x}_i=&~G_{g,i}\left( u_i, x_i \right),\label{DUG dynamics}\\
\varepsilon  \dot{\upsilon}_i=&-\upsilon_i- \mathbb{L}_i\upsilon  -  \mathbb{L}_i \nu  +n\hat{I}_i,
\label{singular perturbation system 1}\\
\varepsilon \dot{\nu}_i=&~\mathbb{L}_i \upsilon,
\label{singular perturbation system 2}\\
\dot{u}_i=&- \alpha_{u_i} u_i +\gamma_i- \epsilon I_{i} \\
\dot{\hat{x}}_i=& -r_i\bar{F}_i \left (\hat{x}_{i},\upsilon_i \right) -r_i A_{i}^\top\lambda _{i}-\gamma _{i} D_i,
\\
\dot{\lambda}_i=&~r_i\normal( A_i\hat{x}_i-s_{A_i}\normal)- r_i \boldsymbol{L}_i A_r \lambda -r_i \boldsymbol{L}_i \theta,
\label{singular perturbation system 4}\\
\dot{\theta}_i=&~\boldsymbol{L}_i A_r\lambda,
\label{singular perturbation system 6}\\
\dot{\gamma}_i=&-u_i+ D_i^\top \hat{x}_i,
\end{align}
\end{subequations}
where \eqref{DUG dynamics} denotes the dynamics of each DGU $i\in \mathcal{N}$, and the non-negative constants $\varepsilon$ and $\epsilon$ denote the control system parameters.
For the sake of the later convergence analysis, let $s_f \triangleq  \col\normal \{ \upsilon, \nu \normal\}$ and $s_d\triangleq \col \normal \{ u, \hat{x}, \lambda, \theta, \gamma \normal\}$.
Then we can write \eqref{whole sysem} as:
\begin{subequations}
\label{compact form of the whole system}
\begin{align}
     \dot{x}=&~G_g \normal( u, x \normal),
    \label{compact slow system 1} \\
    \varepsilon\dot{s}_f=&~G_f\normal( s_f, \hat{I} \normal ),
    \label{compact fast system}\\ 
    \dot{s}_d=&~G_d\normal( \upsilon,s_d, I \normal ). \label{compact slow system 2}
\end{align}
\end{subequations}
where we omits the detailed definitions of the maps $G_g$, $G_f$ and $G_d$.
Note that, in the framework of singular perturbation system theory {\cite{KhalilHK2002}}, \eqref{compact fast system} describes the dynamics of the fast system, while  \eqref{compact slow system 1} and \eqref{compact slow system 2} those of the slow system.
Let $h\normal( \hat{I} \normal)\triangleq\col \normal\{h_{\upsilon}\normal( \hat{I} \normal),h_{\nu}\normal( \hat{I} \normal)\normal\}$ and $s_b\triangleq s_f-h\normal( \hat{I} \normal)$
represent the solution of the equation $G_f\normal( s_f, \hat{I} \normal )=\boldsymbol{0}_{2n}$ and the corresponding boundary layer system state, respectively.  We can write the boundary layer system and reduced-order system as follows:
\begin{subequations}
\begin{align}
\dot{x}=&~ G_g \normal( u, x \normal),
    \label{reduced system 1}\\
    \varepsilon \dot{s}_b=&~G_f\normal( s_b+h, \hat{I}\normal ),\label{boundary layer system}\\
    \dot{s}_d=&~G_d\normal( h_{\upsilon},s_d,I \normal ),
   \label{reduced system 2}
\end{align}
\end{subequations}
where we abbreviate $h\normal( \hat{I} \normal)$ as $h$.

\begin{Theorem}(\textbf{Convergence analysis})\label{Theorem 1}
Let Assumptions \ref{non-empty assumption} and \ref{Monotonicity Assumption} hold and the initial state  $\nu_0$ satisfy $ \boldsymbol{1}_n^\top  \nu_0 =0$. Then, there exists a $\varepsilon^*\in\mathbb{R}_+$ such that \eqref{whole sysem} converges to the largest invariant set $\varPhi_{s,f}$ for all $\varepsilon$ satisfying $0<\varepsilon<\varepsilon^*$, where
\begin{align}
\varPhi_{s,f}=\Biggg\{ \normal( s_f,x, s_d \normal ) \left| \,\, 
\begin{matrix}[1.2]
G_g \normal( u, x \normal)=\boldsymbol{0}\,\,\,\,\,\,\,\,\,
\\
G_f\normal( s_f, \hat{I} \normal )=\boldsymbol{0}\,\,\,\,\,
\\ 
 G_d\normal( \upsilon,s_d, I \normal )=\boldsymbol{0}
\end{matrix} \right. \Biggg\}.
\end{align}
\end{Theorem}

\begin{Proof}
Let $E_b\normal(s_b \normal)$ and $E_s \normal(x, s_d \normal)$ denote respectively the Lyapunov functions of the boundary layer system \eqref{boundary layer system} and the reduced-order system \eqref{reduced system 1}, \eqref{reduced system 2}, i.e., 
\begin{align*}
  E_b \normal(s_b \normal) =&~\sigma \lVert s_b \rVert ^2+\frac{1}{2}\lVert \upsilon_b \rVert ^2+ \frac{1}{2}\lVert \upsilon_b \rVert _{\mathbb{L}}^{2}+\left< \nu_b,\mathbb{L} \upsilon_b  \right>,\\
     E_r\normal(x, s_d  \normal)=&~\frac{1}{2} \big( \lVert \dot{I} \rVert _{L}^{2}+\lVert \dot{I}_l \rVert _{L_l}^{2}+\lVert \dot{V} \rVert _{C}^{2} \big)\\
     &+\frac{1}{2 \epsilon } \|  G_d\normal( h_{\upsilon},s_d, I \normal ) \|^2, 
\end{align*}
where $\sigma$ represents the largest singular value of the Laplacian matrix $\mathbb{L}$.
Then, we can define the composite Lyapunov function as follows:
\begin{align*}
    V\normal( s_b,s_s \normal)= \normal(1-e\normal) E_r\normal(s_s \normal)+e E_b \normal( s_b \normal).
\end{align*}
The convergence analysis  follows  from \cite[Theorem 11.3]{KhalilHK2002}.
For convenience, we define $s_s\triangleq \col\normal\{ x,s_d \normal\}$ and
\begin{align*}
G_s (  h_{\upsilon}, s_s, I \normal )=\left[\,\, 
\begin{matrix}[1.3]
G_g \normal( x,u \normal)
\\
 G_d\normal( h_{\upsilon},s_d, I \normal )
\end{matrix}
\,\,\right].
\end{align*}
Since Proposition \ref{consensus constraint equivalence} ensures that $h_{\upsilon}=\sum_{i=1}^n{\hat{I}_i}$  and it is easy to verify that  $G_d\normal( h_{\upsilon},s_d, I \normal )$ is a monotone function with respect to $s_d$, we can deduce that $J_{G_d,s_d} \normal( h_{\upsilon},s_d, I \normal )$ is positive definite and
\begin{align}
\begin{split}
    \frac{\partial E_r }{ \partial s_d}  G_d\normal( h_{\upsilon},s_d, I \normal ) \le&~\frac{1}{ \epsilon } \normal<G_d\normal( h_{\upsilon},s_d, I \normal ),
    \\
  &~J_{G_d,s_d} \normal( h_{\upsilon},s_d, I \normal ) G_d\normal( h_{\upsilon},s_d, I \normal ) \normal>\\
  &-\normal< \dot{u}, \dot{I} \normal>.
  \end{split}
\end{align}
In addition, we have
\begin{align}
        \frac{\partial E_r }{ \partial x} \dot{x} \le  \dot{x}^\top \left[\,\, \begin{matrix}[1.2]
-R & - \mathbb{I} & \boldsymbol{0} \\
\mathbb{I} & -Z_L^{-1} & B \\
\boldsymbol{0} & -B^\top & -R_l\\
\end{matrix}\,\, \right] \dot{x} +\normal< \dot{u}, \dot{I} \normal>.
\end{align}
Thus we can observe that there exists a positive $\alpha_1$ such that
\begin{align}
    \frac{\partial E_r }{ \partial s_s} G_s\normal( h_{\upsilon}, s_s, I \normal ) \le \alpha_1 \| G_d\normal( h_{\upsilon},s_d, I \normal ) \|^2.
    \label{first condition}
\end{align}
Next we proceed by taking the time derivative of $E_b \normal( s_b \normal)$, and based on the fact that $ \boldsymbol{1}_n^\top s_b =0$ (following from $ \boldsymbol{1}_n^\top  \nu_0 =0$), we have:
\begin{align*}
    &\frac{\partial E_b }{ \partial s_b} \dot{s}_b
    =\\
    &-s_b^\top \underset{\triangleq~A_{\alpha}}{\underbrace{ \left[ \,\,\begin{matrix}[1.2]
	\left( 2\sigma +1 \right) \mathbb{I}+\left( 2\sigma +2 \right) \mathbb{L}
 &		\mathbb{L}+\mathbb{L}^2\\
	\mathbb{L}+\mathbb{L}^2&		\mathbb{L}^2+\alpha_{\nu} \boldsymbol{1}_{n \times n}\\
\end{matrix} \,\, \right]}} s_b,
\end{align*}
where $\alpha_{\nu}$ is a constant.
Then, one can verify that all the eigenvalues of the matrix $A_{\alpha}$ are positive (we omit the proof due to space limitation).
Hence, based on the property of the Rayleigh quotient, we have
\begin{align}
    \frac{\partial E_b }{ \partial s_b} \dot{s}_b\le \delta_{\min} \normal(A_{\alpha}\normal)\|s_b \|^2=\alpha_2  \|s_b \|^2.
    \label{second condition}
\end{align}
Now, since the function $G_s \normal (  s_b+h_{\upsilon}, s_s, I \normal )$ is linear with respect to $s_b+h_{\upsilon}$, we can deduce that there exists a positive constant $\beta_1$ such that
\begin{align}
\begin{split}
   \frac{\partial E_r }{ \partial s_s}\normal[  G_s (  s_b+h_{\upsilon}, s_s, I \normal )&-G_s (  h_{\upsilon}, s_s, I \normal ) \normal]
   \\
   \le&~\beta_1  \| G_d\normal( h_{\upsilon},s_d, I \normal)  \| \| s_b \|.
   \end{split}
   \label{third condition} 
\end{align}
\end{Proof}
Finally, one can show that \eqref{the third condition} holds.
\begin{figure*}
    \centering
\begin{align}
\begin{split}
    \bigg[ \frac{\partial E_b }{ \partial s_s}-\frac{\partial E_b }{ \partial s_b}\frac{\partial h }{ \partial s_s} \bigg] G_s (  s_b+h_{\upsilon}, s_s, I \normal )
    \le   s_b^\top \left[\,\, \begin{matrix}
\sigma\mathbb{I}+ \mathbb{I}+\mathbb{L} & \mathbb{L} \\
\mathbb{L} & \sigma\mathbb{I}
    \end{matrix}\,\, \right] \Bigg( \frac{\partial h }{ \partial s_s}  G_s (  h_{\upsilon}, s_s, I \normal )-\frac{\partial h }{ \partial \hat{I}}  \normal(p_r V^r \circ s_b  \normal) \Bigg).
\end{split}
    \label{the third condition}
\end{align}
\end{figure*}
Furthermore, since $\partial h / \partial s_s$ and $\partial h / \partial \hat{I}$ are constant matrices, then we can deduce that there exist two positive constants $\beta_2$ and $\xi$ such that the following inequality holds
\begin{align}
\begin{split}
    \bigg[ \frac{\partial E_b }{ \partial s_s}-\frac{\partial E_b }{ \partial s_b}\frac{\partial h }{ \partial s_s} &\bigg] G_s (  s_b+h_{\upsilon}, s_s, I \normal )
    \\
    \le &~\beta_2 \| G_d\normal( h_{\upsilon},s_d, I \normal)  \| \| s_b \| +\xi \|s_b\|^2.
    \end{split}
    \label{fourth condition}
\end{align}
So far, we have verified all the conditions in \cite[Theorem 11.3]{KhalilHK2002}, i.e,  \eqref{first condition}, \eqref{second condition}, \eqref{third condition} and \eqref{fourth condition}.
Hence, we can conclude that if 
\begin{align*}
0<\varepsilon \le \varepsilon _{e}^{*}=\frac{\alpha _1\alpha _2}{\alpha_1 \xi +\beta _1\beta _2},
\end{align*}
then system \eqref{whole sysem} converges to the largest invariant \mbox{set $\varPhi_{s,f}$.
\hfill\QED}
\section{Simulations}
\label{simulcation}

In this section, we assess the performance of the proposed distributed control system \eqref{whole sysem} in simulation, considering a microgrid with four DGUs in a ring topology.
We set the price parameters $p_r$ and $l $ as $5$ and $0.01$, respectively. Also, we select the fast system parameter $\varepsilon$ equal to $0.01$.
The parameters of the objective and penalty functions are reported in Table. \ref{Parameters setting for the objective functions and the penalty functions}.
We report the parameters of all the DGUs and transmission lines in Tables \ref{Parameters setting for all DGUs and loads} and \ref{Parameters setting for the transmission lines}.

\begin{table}
\scriptsize
\centering
\caption{Parameters of the objective and penalty functions.}
\label{Parameters setting for the objective functions and the penalty functions}
\begin{tabular}{cccccccc}
\toprule
\toprule
Number & $r_i$  & $\alpha_{I_i}$ & $\alpha_{V_i}$ & $\alpha_{u_i}$ & $\alpha_{I_{l,k}}$ &$\rho_{V_i}$ &$\rho_{I_{l,k}}$ \\
\midrule
\midrule
1      & 1.0060 & 10.6569          & 0.7516      & 1.0155       & 1.3724  &1200  & 1000       \\
2      & 1.0399 & 10.6280          & 0.6203      & 1.9841       & 1.1981 &1200  & 1000        \\
3      & 1.0527 & 10.2920          & 0.8527      & 1.1672       & 1.4897 &1200  & 1000        \\
4      & 1.0417 & 10.4317          & 0.9379      & 1.1060       & 1.3395  &1200  & 1000 \\     
\bottomrule
\bottomrule 
\end{tabular}
\end{table}

\begin{table*}
\footnotesize
\centering
\caption{Parameters of the DGUs and loads}
\label{Parameters setting for all DGUs and loads}
\begin{tabular}{ccccccccccc}
\toprule
\toprule
DGU Number & $L_i$ (mH) & $C_i$ (mF) & $R_i$ (m$\Omega$) & $I_i^r$ (A) & $u_i^r$ (V) & $V_i^r$ (V) &$V_i^{\min}$ (V) &$V_i^{\max}$ (V)& $Z_{L,i}$ ($\Omega$) & $I_{L,i}$ (A)  \\ 
\midrule
\midrule
1       & 1.8      & 2.2    & 20         &0      &0      & 380  &377&383 & 16   & 30         \\
2       & 2.0      & 1.9    & 18         &0      &0      & 380   &377&383& 50   & 15                    \\
3       & 3.0      & 2.5    & 16         &0      &0      & 380   &377&383& 16   & 30                    \\ 
4       & 2.2      & 1.7    & 15         &0      &0      & 380   &377&383& 20   & 26         \\ 
\bottomrule
\bottomrule
\end{tabular}
\end{table*}

\begin{figure}
\includegraphics[width=8cm]{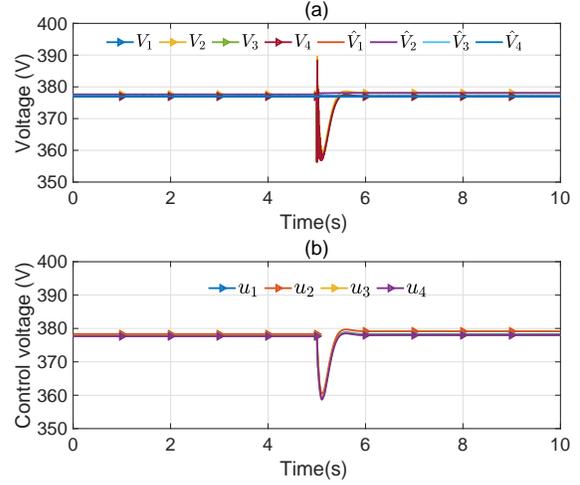}
\caption{(a) The loads' and decision system' voltages. (b) The DGU control voltages. }
\label{Fig 1}
\end{figure}

\begin{figure}
\includegraphics[width=8cm]{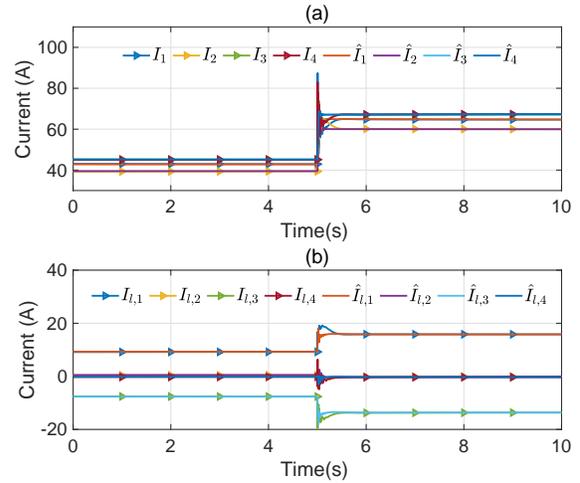}
\caption{  (a) The DGUs' and the decision system's currents. (b) The transmission lines' and decision system's currents.}
\label{Fig 2}
\end{figure}

\begin{table*}
\footnotesize
\centering
\caption{Parameters of the transmission lines.}
\label{Parameters setting for the transmission lines}
\begin{tabular}{ccccccccc}
\toprule
\toprule
Line Number & Head node & Tail node & $R_{l,k}$ (m$\Omega$) & $L_{l,k}$ ($\mu$H) & $I_{l,k}^{\min}$ (A) & $I_{l,k}^{\max}$ (A) & $I_{l,k}^r$ (A) & Manage agent \\
\midrule
\midrule
1      & 1         & 2         & 70                & 2.1            & -20          & 20           &0     &1             \\
2      & 2         & 3         & 50                & 2.0            & -20          & 20          &0     &2 \\
3      & 3         & 4         & 80                & 3.0            & -20          & 20          &0     &3 \\
4      & 4         & 1         & 60                & 2.2            & -20          & 20          &0     &1           
\\
\bottomrule
\bottomrule
\end{tabular}
\end{table*}
We consider that the microgrid initial conditions are within the feasible set and the system remain unperturbed for the first $5$ seconds. Then, at the time instant $t = 5$ s, each current-type load $I_{L,i}$ and resistance-type load $Z_{L,i}$ is decreased by 3 units.
We present the results in Fig. \ref{Fig 1} and \ref{Fig 2}, and we observe that the microgrid's states  converge to the equilibrium within a short time after the loads change.
Moreover, the new equilibrium satisfies the optimal condition~\eqref{Final KKT condition}.

In general, the simulation results show excellent performance both in terms of optimality and transient response.



\section{conclusion}
\label{conclusion}
In this paper, we first design a distributed control system to solve the power trading problem (described as an aggregative game) in a DC microgrid.
Then, we interconnect the designed control  system with the microgrid in a passive way and analyze the convergence of the overall closed-loop system.
Although we prove that there exists $\varepsilon^*$ for the fast system, we do not explicitly provide its exact bound, which is left as a future work.

\bibliographystyle{IEEEtran}

\bibliography{IEEEabrv,mylib}

\begin{thebibliography}{10}
\providecommand{\url}[1]{#1}
\csname url@samestyle\endcsname
\providecommand{\newblock}{\relax}
\providecommand{\bibinfo}[2]{#2}
\providecommand{\BIBentrySTDinterwordspacing}{\spaceskip=0pt\relax}
\providecommand{\BIBentryALTinterwordstretchfactor}{4}
\providecommand{\BIBentryALTinterwordspacing}{\spaceskip=\fontdimen2\font plus
\BIBentryALTinterwordstretchfactor\fontdimen3\font minus
  \fontdimen4\font\relax}
\providecommand{\BIBforeignlanguage}[2]{{%
\expandafter\ifx\csname l@#1\endcsname\relax
\typeout{** WARNING: IEEEtran.bst: No hyphenation pattern has been}%
\typeout{** loaded for the language `#1'. Using the pattern for}%
\typeout{** the default language instead.}%
\else
\language=\csname l@#1\endcsname
\fi
#2}}
\providecommand{\BIBdecl}{\relax}
\BIBdecl

\bibitem{cenedese:2019:PEV_charging}
C.~Cenedese, F.~Fabiani, M.~Cucuzzella, J.~M. Scherpen, M.~Cao, and
  S.~Grammatico, ``Charging plug-in electric vehicles as a mixed-integer
  aggregative game,'' in \emph{2019 IEEE 58th Conference on Decision and
  Control (CDC)}.\hskip 1em plus 0.5em minus 0.4em\relax IEEE, 2019, pp.
  4904--4909.

\bibitem{YuW2017}
W.~Yu, C.~Li, X.~Yu, G.~Wen, and J.~L{\"u}, ``Economic power dispatch in smart
  grids: a framework for distributed optimization and consensus dynamics,''
  \emph{Science China Information Sciences}, vol.~61, no.~1, pp. 1--16, 2018.

\bibitem{CucuzzellaM2021}
M.~Cucuzzella, T.~Bouman, K.~Kosaraju, G.~Schuitema, N.~H. Lemmen,
  S.~Johnson~Zawadzki, C.~Fischione, L.~Steg, and J.~M. Scherpen, ``Distributed
  control of dc grids: integrating prosumers motives,'' \emph{IEEE Transactions
  on Power Systems}, pp. 1--1, 2021.

\bibitem{CucuzzellaM2018}
M.~Cucuzzella, S.~Trip, C.~De~Persis, X.~Cheng, A.~Ferrara, and A.~van~der
  Schaft, ``A robust consensus algorithm for current sharing and voltage
  regulation in dc microgrids,'' \emph{IEEE Transactions on Control Systems
  Technology}, vol.~27, no.~4, pp. 1583--1595, 2019.

\bibitem{KosarajuKC2021}
K.~C. Kosaraju, M.~Cucuzzella, J.~M.~A. Scherpen, and R.~Pasumarthy,
  ``Differentiation and passivity for control of brayton–moser systems,''
  \emph{IEEE Transactions on Automatic Control}, vol.~66, no.~3, pp.
  1087--1101, 2021.

\bibitem{cenedese:2021:ad_geno}
C.~Cenedese, G.~Belgioioso, S.~Grammatico, and M.~Cao, ``An asynchronous
  distributed and scalable generalized {N}ash equilibrium seeking algorithm for
  strongly monotone games,'' \emph{European Journal of Control}, vol.~58, pp.
  143--151, 2021.

\bibitem{cenedese:2021:tv_prox_dyn}
C.~Cenedese, G.~Belgioioso, Y.~Kawano, S.~Grammatico, and M.~Cao,
  ``Asynchronous and {T}ime-{V}arying {P}roximal {T}ype {D}ynamics in
  {M}ultiagent {N}etwork {G}ames,'' \emph{IEEE Transactions on Automatic
  Control}, vol.~66, no.~6, pp. 2861--2867, 2021.

\bibitem{Dorfler2017}
F.~Dörfler, S.~Bolognani, J.~W. Simpson-Porco, and S.~Grammatico,
  ``Distributed control and optimization for autonomous power grids,'' in
  \emph{2019 18th European Control Conference (ECC)}, 2019, pp. 2436--2453.

\bibitem{DePersisC2019}
C.~De~Persis and S.~Grammatico, ``Continuous-time integral dynamics for a class
  of aggregative games with coupling constraints,'' \emph{IEEE Transactions on
  Automatic Control}, vol.~65, no.~5, pp. 2171--2176, 2020.

\bibitem{FacchineiF2014}
F.~Facchinei, J.-S. Pang, G.~Scutari, and L.~Lampariello, ``Vi-constrained
  hemivariational inequalities: distributed algorithms and power control in
  ad-hoc networks,'' \emph{Mathematical Programming}, vol. 145, no.~1, pp.
  59--96, 2014.

\bibitem{BelgioiosoG2018}
G.~Belgioioso and S.~Grammatico, ``Projected-gradient algorithms for
  generalized equilibrium seeking in aggregative games arepreconditioned
  forward-backward methods,'' in \emph{2018 European Control Conference (ECC)},
  2018, pp. 2188--2193.

\bibitem{boyd2004}
S.~Boyd, S.~P. Boyd, and L.~Vandenberghe, \emph{Convex optimization}.\hskip 1em
  plus 0.5em minus 0.4em\relax Cambridge university press, 2004.

\bibitem{YiP2019}
P.~Yi and L.~Pavel, ``An operator splitting approach for distributed
  generalized nash equilibria computation,'' \emph{Automatica}, vol. 102, pp.
  111--121, 2019.

\bibitem{LiangS2017}
S.~Liang, P.~Yi, and Y.~Hong, ``Distributed nash equilibrium seeking for
  aggregative games with coupled constraints,'' \emph{Automatica}, vol.~85, pp.
  179--185, 2017.

\bibitem{Van2000}
A.~Van~der Schaft, \emph{L2-gain and passivity techniques in nonlinear
  control}.\hskip 1em plus 0.5em minus 0.4em\relax Springer, 2000.

\bibitem{Cherukuri2015}
A.~Cherukuri and J.~Cortés, ``Distributed generator coordination for
  initialization and anytime optimization in economic dispatch,'' \emph{IEEE
  Transactions on Control of Network Systems}, vol.~2, no.~3, pp. 226--237,
  2015.

\bibitem{NabetaniK2021}
K.~Nabetani, P.~Tseng, and M.~Fukushima, ``Parametrized variational inequality
  approaches to generalized nash equilibrium problems with shared
  constraints,'' \emph{Computational Optimization and Applications}, vol.~48,
  no.~3, pp. 423--452, 2011.

\bibitem{cenedese:2020:time_var_constrained}
C.~Cenedese, G.~Belgioioso, S.~Grammatico, and M.~Cao, ``Time-varying
  constrained proximal type dynamics in multi-agent network games,'' in
  \emph{2020 European Control Conference (ECC)}, 2020, pp. 148--153.

\bibitem{FacchineiF2003}
F.~Facchinei and J.-S. Pang, \emph{Finite-dimensional variational inequalities
  and complementarity problems}.\hskip 1em plus 0.5em minus 0.4em\relax
  Springer, 2003.

\bibitem{HanSP1979}
S.-P. Han and O.~L. Mangasarian, ``Exact penalty functions in nonlinear
  programming,'' \emph{Mathematical programming}, vol.~17, no.~1, pp. 251--269,
  1979.

\bibitem{KhalilHK2002}
H.~K. Khalil, ``Nonlinear systems third edition,'' \emph{Patience Hall}, vol.
  115, 2002.

\end{thebibliography}

\end{document}